\documentclass[pra,twocolumn,nofootinbib,floatfix,10pt]{revtex4-2}

\usepackage{amsmath}
\usepackage{amssymb}
\usepackage{wasysym}
\usepackage{graphicx}
\usepackage{color,soul}
\usepackage{physics}
\usepackage{siunitx}
\usepackage{dsfont}
\usepackage{float}
\usepackage{xcolor}
\usepackage[english]{babel}
\usepackage{blindtext}
\usepackage[english,nomargin,inline,marginclue,draft]{fixme}
\pdfpageheight\paperheight
\pdfpagewidth\paperwidth

\usepackage[colorlinks,linkcolor=blue,anchorcolor=blue,citecolor=blue,urlcolor=blue]{hyperref}

\fxusetheme{colorsig}
\FXRegisterAuthor{cg}{acg}{CG}  
\FXRegisterAuthor{th}{ath}{\color{blue}TH}  
\FXRegisterAuthor{ib}{aib}{\color{red}IB} 
\FXRegisterAuthor{sh}{ash}{\color{cyan}SH} 
\FXRegisterAuthor{db}{adb}{\color{green}DB} % now I can ..
\FXRegisterAuthor{ps}{aps}{PS}
\makeatletter
\renewcommand*\FXLayoutInline[3]{%
  {\@fxuseface{inline}\ignorespaces{\color{fx#1}[#3: #2]}}}
\makeatother

\long\def\symbolfootnote[#1]#2{\begingroup%
\def\thefootnote{\fnsymbol{footnote}}\footnotetext[#1]{#2}\endgroup}

\def\nobreakbefore{%
  \relax\ifvmode\else
    \ifhmode
      \ifdim\lastskip > 0pt\relax
        \unskip\nobreakspace
      \else % added to put a ~if no space was typed. (Unclear why it sometimes worked before )
        \nobreakspace
      \fi
    \fi
  \fi
}
\let\oldcite\cite
\renewcommand\cite{\nobreakbefore\oldcite}

\begin{document}
\title{Observation of non-Hermitian topology in cold Rydberg quantum gases}

\author{Jun Zhang$^{1,2,\textcolor{blue}{\dagger}}$}
\author{Ya-Jun Wang$^{1,2,\textcolor{blue}{\dagger}}$}
\author{Shi-Yao Shao$^{1,2,\textcolor{blue}{\dagger}}$}
\author{Bang Liu$^{1,2}$}
\author{Li-Hua Zhang$^{1,2}$}
\author{Zheng-Yuan Zhang$^{1,2}$}
\author{Xin Liu$^{1,2}$}
\author{Chao Yu$^{1,2}$}
\author{Qing Li$^{1,2}$}
\author{Han-Chao Chen$^{1,2}$}
\author{Yu Ma$^{1,2}$}
\author{Tian-Yu Han$^{1,2}$}
\author{Qi-Feng Wang$^{1,2}$}
\author{Jia-Dou Nan$^{1,2}$}
\author{Yi-Ming Yin$^{1,2}$}
\author{Dong-Yang Zhu$^{1,2}$}
\author{Qiao-Qiao Fang$^{1,2}$}
\author{Dong-Sheng Ding$^{1,2,\textcolor{blue}{\star}}$}
\author{Bao-Sen Shi$^{1,2}$}

\affiliation{$^1$Key Laboratory of Quantum Information, University of Science and Technology of China, Hefei, Anhui 230026, China.}
\affiliation{$^2$Synergetic Innovation Center of Quantum Information and Quantum Physics, University of Science and Technology of China, Hefei, Anhui 230026, China.}

\date{\today}

\symbolfootnote[1]{J.Z, Y.J.W and S.Y.S contribute equally to this work.}
\symbolfootnote[2]{dds@ustc.edu.cn}

\maketitle
\textbf{The pursuit of topological phenomena in non-Hermitian systems has unveiled new physics beyond the conventional Hermitian paradigm, yet their realization in interacting many-body platforms remains a critical challenge. Exploring this interplay is essential to understand how strong interactions and dissipation collectively shape topological phases in open quantum systems. Here, we experimentally demonstrate non-Hermitian spectra topology in a dissipative Rydberg atomic gas and characterize parameters-dependent winding numbers. By increasing the interaction strength, the system evolves from Hermitian to non-Hermitian regime, accompanying emergence of trajectory loop in the complex energy plane. As the scanning time is varied, the spectra topology becomes twisted in the complex energy plane manifesting as a topology phase transition with the sign winding number changed. When preparing the system in different initial states, we can access a nontrivial fractional phase within a parameter space that globally possesses an integer winding. Furthermore, by changing the scanning direction, we observe the differentiated loops, revealing the breaking of chirality symmetry. This work establishes cold Rydberg gases as a versatile platform for exploring the rich interplay between non-Hermitian topology, strong interactions, and dissipative quantum dynamics.}

The phase in topology reveals the inherent space property that is preserved under continuous deformations, such as stretching and bending. Topological phase transitions refer to changes in the state of matter that are characterized by alterations in the topological properties of the system, rather than conventional symmetry-breaking mechanisms \cite{xu2011topological,caraglio2015stretching,kawabata2019topological,kawabata2019symmetry,ji2020categorical,xia2021nonlinear}. Studying topological phases and their transitions involves exploring materials' features with distinct global geometric features and investigating how these phases change under different conditions \cite{xu2012observation,heide2022probing,ding2022non,lin2022experimental}. These transitions can occur between different topological phases, where the properties of the system remain invariant under continuous transformations but differ in their topological invariants\cite{zhu2013topological,mittal2016measurement,lee2019topological,wang2021detecting,li2023direct,liu2025zak}, such as the presence of edge states\cite{xiao2017observation,de2019observation,cai2021topological,yang2022topological,hao2023topological,yang2025observing,yue2025observing}, topological insulators \cite{insulator2011topological,zhang2013topology,lin2022observation,lin2022observation} and superconductors \cite{zhang2013majorana,zhang2018observation} in quantum systems. In addition, the topological phase transitions are also investigated in a non-equilibrium system \cite{vajna2015topological,heyl2017dynamical}, non-Hermitian system \cite{gong2018topological,longhi2019topological,lin2022topological,dai2024non,li2024observation}, and dynamics systems \cite{flaschner2018observation,heyl2018dynamical,valencia2024rydberg,xiao2024dynamical}, in which dynamical behaviors and interactions enrich topological properties.

Due to the exaggerated properties of Rydberg atoms \cite{saffman2010quantum,adams2019rydberg,browaeys2020many}, they not only enhance the complexity of the system's behavior but also provides a platform for investigating topological states and their phase transitions, for example, topological band structure \cite{de2019observation} and topological order in the Kagome lattice \cite{samajdar2021quantum}, and other exotic topological features \cite{li2015exotic,li2021symmetry,verresen2021prediction,kanungo2022realizing}. The characteristic of long-range interaction makes the laser-driven Rydberg atomic system displaying rich non-equilibrium physics and nonlinear dynamics \cite{lee2012collective,carr2013nonequilibrium,schempp2014full,marcuzzi2014universal,lesanovsky2014out,urvoy2015strongly,gambetta2019,ding2022enhanced, Signatures2020Helmrich,ding2019Phase,Wintermantel2020Cellular, klocke2021hydrodynamic, Wadenpfuhl2023Synchronization, ding2023ergodicity, wu2024dissipative,liu2024higher,liu2024microwave,liu2024bifurcation}. The interactions in Rydberg atoms induce non-Hermitian properties \cite{delplace2021symmetry} and hysteresis loops characterized by asymmetrical responses to varying probe intensities \cite{zhang2025exceptional}. The precise control over Rydberg atom excitation facilitates the creation of effective non-Hermitian Hamiltonians, which exhibit rich dynamical evolution. This capability provides a powerful platform for investigating the emergence of dynamical topological phases and the symmetry breaking.

\begin{figure*}
\centering
\includegraphics[width=0.95\linewidth]{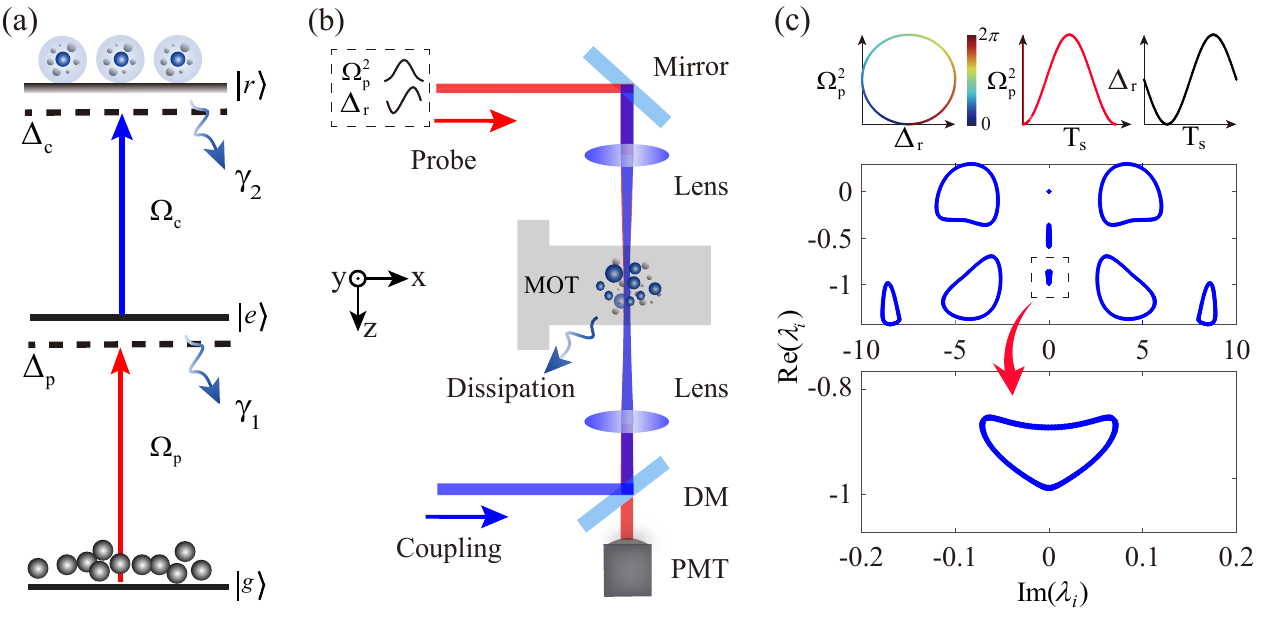}
\caption{\textbf{Experimental diagram and energy spectral topology.} (a) Energy level diagrams. Probe and coupling fields excite atoms from the ground state $\ket{g}$ to the Rydberg state $\ket{r}$. (b) Schematic diagram of the experimental setup. The probe beam is incident opposite to the coupling beam and focused in cold $\rm{^{85}Rb}$ atoms. MOT: magneto-optical trap, PMT: photomultiplier tube, DM: dichroic mirror. (c) Numerical simulations of Liouvillian eigenspectrum varying with parameter $\rm{Arg(Z)}$ on the complex energy plane, where $\text{Z}=\rm{\Delta_r}+i\rm{\Omega_{p}^2}$. The parameters are $\{ \rm{\Omega_c},\rm{\Delta _c},\gamma_{\rm{eff}},\gamma_1,\gamma_2 \}$ = $\{4,0,2,0.8,1.2\}$, $\rm{\Omega_p^2 }\in[0,6]$, and $\rm{\Delta _r} \in[-3,3]$. The enlarged view represents the eigenvalue  $\lambda_{rr}$ of eigenstate $\rho_{rr}$. The upper panel shows the diagram of changing trends of parameters $\rm{\Omega^2 _p}$ and $\rm{\Delta _r}$.}
\label{Fig1}
\end{figure*}

In this work, we experimentally demonstrate non-Hermitian spectral topology in a dissipative cold Rydberg atomic gas. The emergence of a distinct spectral loop in the complex energy plane reveals a transition from a Hermitian to a non-Hermitian regime. Through precise measurements of the parameter-dependent complex energy spectra, we extract the corresponding topological winding numbers. Remarkably, we find that the topology itself is not static but dynamical. By tuning the parameter scanning time, the spectral loop can be “twisted”, leading to a topological phase transition accompanied by a sign change in the winding number. Then, by preparing the system in different initial states, we access a nontrivial fractional phase within a spectrally topological loop. Furthermore, by reversing the parameter scan direction, we observe a differentiation in the resulting spectral loops. This breaking of chiral symmetry in the measurement trajectory provides non-trivial behavior in the non-Hermitian many-body system. The reported results show that the strongly interacting cold Rydberg gases can be regarded as a versatile platform for exploring non-Hermitian topological physics.

\subsection*{Physical model}
To investigate the energy spectrum topological properties, we consider an interacting three-level Rydberg atomic system as illustrated in Fig.~\ref{Fig1}(a). The system consists of three atomic state manifolds: the ground state $\ket{g}$, the intermediate excited state $\ket{e}$, and the Rydberg state $\ket{r}$. The probe field with Rabi frequency (detuning) $\rm{\Omega_{p}}$ ($\rm{\Delta_{p}}$) drives the transition $\ket{g}\leftrightarrow\ket{e}$. The coupling field, with Rabi frequency $\rm{\Omega_{c}}$ and detuning $\rm{\Delta_{c}}$, couples the transition $\ket{e}\leftrightarrow\ket{r}$. The spontaneous decay rates for the states $\ket{e}$ and $\ket{r}$ are $\gamma_1$ and $\gamma_2$, respectively. The experimental setup is depicted by Fig.~\ref{Fig1}(b). 

To explore the topological properties of the system, we first analyze its Hamiltonian. The effective single-particle Hamiltonian in the interaction picture and rotating-wave approximation can be written as:
\begin{equation}
\begin{aligned}
H_{\rm{eff}}=&\sum\limits_j {[\frac{{{\rm{\Omega _p}}}}{2}\left| e \right\rangle {{\left\langle g \right|}_j} + \frac{{{\rm{\Omega _c}}}}{2}\left| r \right\rangle {{\left\langle e \right|}_j}+ H.c.]}\cr
&- \sum\limits_j {[{\rm{\Delta _p}}\left| e \right\rangle {{\left\langle e \right|}_j}+ ({\rm{\Delta _r}+\mathit{i} \gamma_{\rm{eff}})}\left| r \right\rangle {{\left\langle r \right|}_j}}],
\end{aligned}
\end{equation}
where $\rm{\Delta _r}=\rm{\Delta _p}+\rm{\Delta _c}$ represents two-photon detuning. The interactions between atoms in the states $\left| r \right\rangle$ influence the many-body quantum dynamics. Here, the interactions induce an additional dissipation on the state $\left| r \right\rangle$, resulting in a dissipation term $i\gamma_{\rm{eff}}$. 

\begin{figure*}
\centering
\includegraphics[width=1\linewidth]{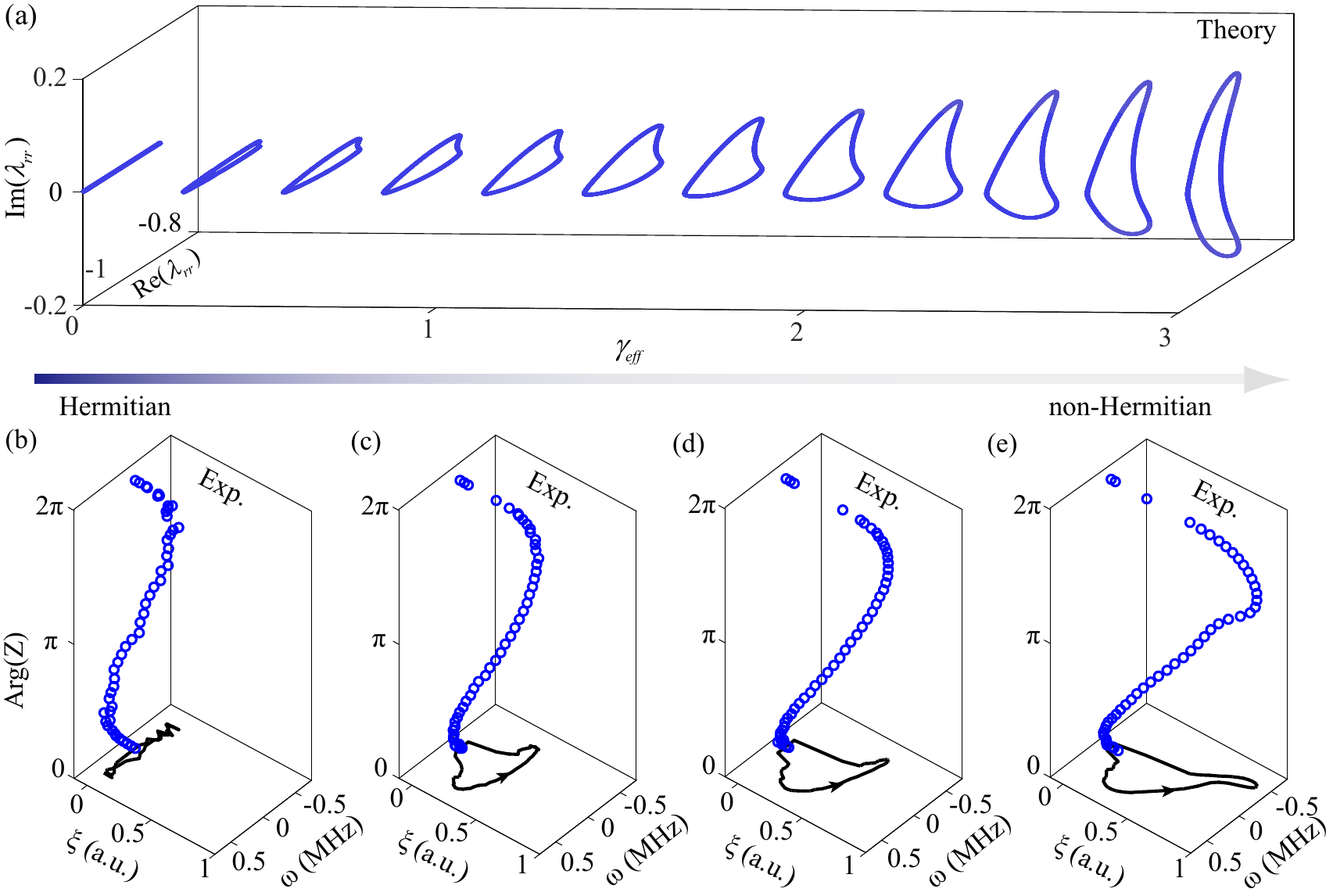}
\caption{\textbf{Non-Hermitian energy spectral topology.} (a) Numerical simulation of the energy eigenvalue $\lambda_{rr}$ of Rydberg steady state versus $\rm{\gamma_{eff}}$ on the complex plane, where the parameters are $\{ \rm{\Omega_c},\rm{\Delta _c},\gamma_1,\gamma_2 \}$ = $\{4,0,0.8,1.2\}.$ The Liouvillian eigenspectrum forms a loop in the complex plane. (b-e) The evolution trajectory of Rydberg eigenstate in the complex energy plane is strongly correlated with the variation of probe transmission and the energy level detuning $\rm{\Delta_r}$ in the experiment varies with the parameter Arg(Z), where $\text{Z}=\rm{\Delta_r}+\mathit{i}\rm{\Omega^2_{p}}$ in the scanning time of $\rm{T_s}=18~\rm{\mu s}$. Panels (b), (c), (d), and (e) display the spectral features for $(\rm{\Omega_p}/2\pi)^2\in[0,1]~\rm{MHz}^2$, $(\rm{\Omega_p}/2\pi)^2\in[0,4]~\rm{MHz}^2$, $(\rm{\Omega_p}/2\pi)^2\in[0,10]~\rm{MHz}^2$,  and $(\rm{\Omega_p}/2\pi)^2\in[0,20]~\rm{MHz}^2$ , respectively.}
\label{Fig2}
\end{figure*}

Under the mean field approximation (the interaction between Rydberg atoms is equivalent to the effect of a single atom in the collective field of surrounding atoms), $\gamma_{\rm{eff}}$ is proportional to the population of $\ket{r}$ state~\cite{zhang2025exceptional}. The Lindblad master equation including the associated dissipative terms can be written as
\begin{equation}
\dot{\rho} = -i[H_\mathrm{eff}, \rho] + \sum_{j} \left( L_j \rho L_j^{\dagger} - \frac{1}{2} \left\{ L_j^{\dagger} L_j, \rho \right\} \right) = \mathcal{L}\rho,
\end{equation}
where $\mathcal{L}$ denotes the Liouvillian superoperator. To fully investigate the emergence of the system's non-Hermitian topological property, we perform a full spectral decomposition by solving the Liouvillian eigenvalue equation $\mathcal{L} \hat{\rho}_i = \lambda_i \hat{\rho}_i$, $\lambda_i$ and $\hat{\rho}_i$ are the eigenvalues and eigenstates of the Liouvillian, respectively. The results of diagonalizing the system's effect Liouvillian matrix are shown in Fig.~\ref{Fig1}(c), where the trajectory of eigenvalues on the complex plane reveals a `line-gap’ topology, see the Liouvillian eigenspectrum on the complex plane.

We then characterize the spectral topology for our system \cite{ding2022non}, where the Liouvillian space $\mathcal{L}(\rm{Z})$ depends on the complex parameter $\rm{Z}=\rm{\Delta_r}+\mathit{i}\rm{\Omega_p^2}$. Similar to the general non-Hermitian case, the eigenvalues $\lambda(\rm{Z})$$  \in \mathbb{C}$ lie in the complex energy plane.  As $\rm{Z}$ varies along a closed contour $C_\mathrm{Z}$ in the parameter space, the image $C_\mathrm{Z}$ forms a loop that may enclose a finite area in the complex plane. This loop can wind around a reference energy point $\lambda$, giving rise to a nontrivial spectral winding number $\mathcal{W}_\lambda$, which is defined as \cite{ding2022non,budich2020non,wang2021generating}: 
\begin{align}
\mathcal{W}_\lambda=\frac{1}{2 \pi i} \oint_{C_\mathrm{Z}} \mathrm{~d} \vec{\mathrm{Z}} \cdot \nabla_\mathrm{Z} \ln \operatorname{det}\left(\mathcal{L}(\mathrm{Z})-\mathrm{\lambda I}\right).
\end{align}
It counts the number of times the spectrum winds around the reference eigenvalues $\lambda$ and I is an identity matrix of the same dimension as $\mathcal{L}$. For a single clockwise (CW) encirclement, the winding number yields $\mathcal{W}_\lambda = -1$, while a single counterclockwise (CCW) one gives $\mathcal{W}_\lambda = 1$. This quantized winding number reflects the intrinsic spectral topology of the system and is robust against continuous deformations of the contour $C_\mathrm{Z}$, as long as no singularities are crossed. 

\subsection*{Spectral topology on the complex plane}
\begin{figure*}
\centering
\includegraphics[width=1\linewidth]{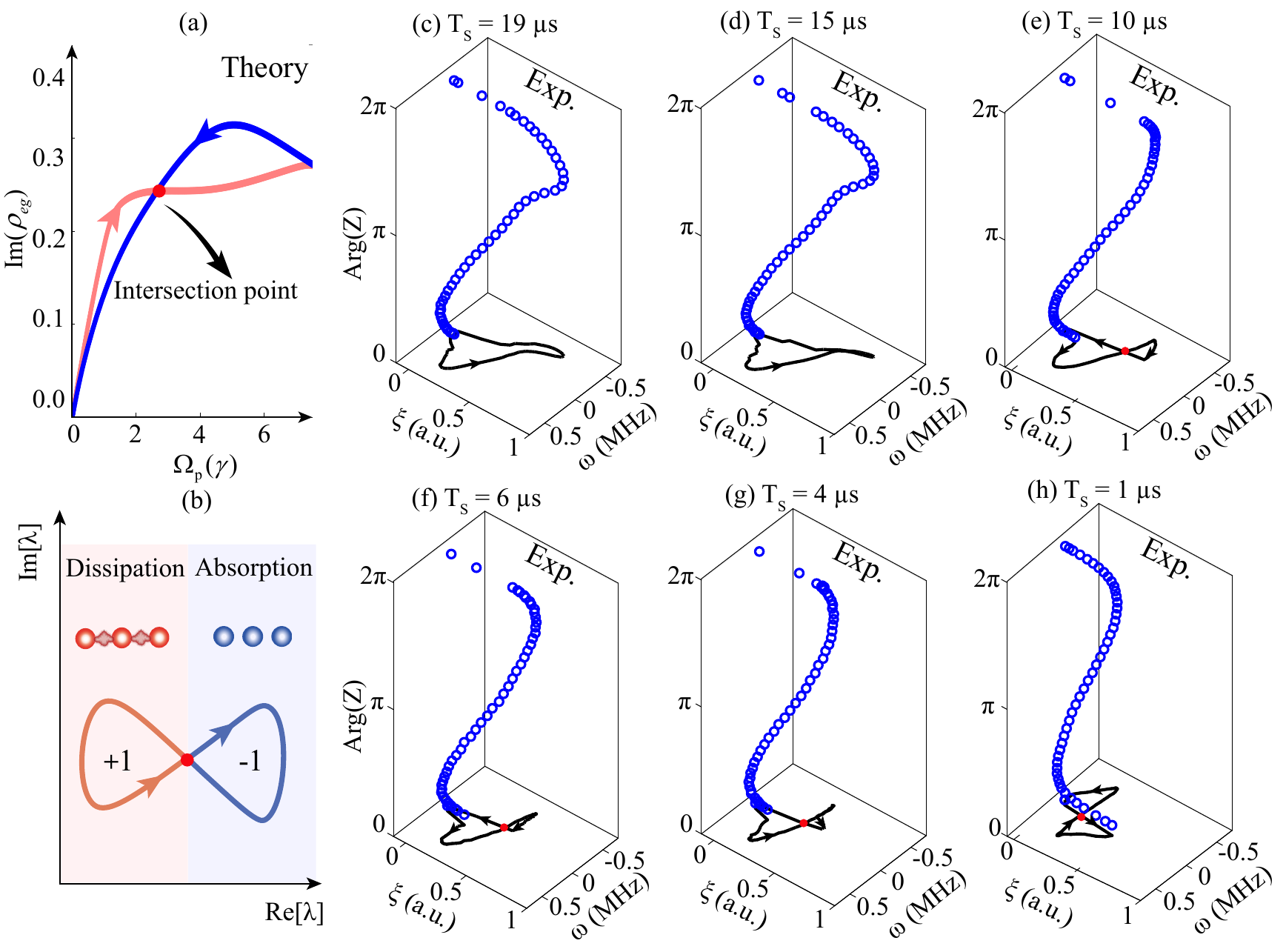}
\caption{\textbf{Twisted topology.} (a) The simulated $\rm{Im(\rho_{eg})}$ during the up-scan and down-scan processes of $\rm{\Omega_p^2}$ through Lindblad master equation. (b) Schematic diagram of the different winding numbers corresponding to the absorption- and dissipation-dominated spectra of the system. (c)-(h) The spectral trajectory of the transmitted signal and the energy level detuning $\rm{\Delta_r}$ in the experiment varies with the parameter Arg(Z). The scanning times are as follows: $\rm{T_s}=19~\rm{\mu s}$ in (c), $\rm{T_s}=15~\rm{\mu s}$ in (d), $\rm{T_s}=10~\rm{\mu s}$ in (e), $\rm{T_s}=6~\rm{\mu s}$ in (f), $\rm{T_s}=4~\rm{\mu s}$ in (g) and $\rm{T_s}=1~\rm{\mu s}$ in (h).}
\label{Fig3}
\end{figure*}

Figure~\ref{Fig2}(${\rm a}$) shows a numerical simulation of the complex energy eigenvalue $\rm{\lambda_{rr}}$ as a function of the dissipation $\gamma_{\rm{eff}}$. As the interaction $\gamma_{\rm{eff}}$ increases, the eigenvalue trajectory evolves into distinct loops on the complex plane. To quantitatively probe the topological features, we employ electromagnetically induced transparency (EIT) in a cold atomic ensemble of $\rm{^{85}Rb}$  atoms. Using an acousto-optic modulator, we periodically sweep the probe field intensity and detuning $\rm{\Delta_r}$ with controlled period $\rm{T_s}$, tracing closed contours in the ($\rm{\Omega^2_p}$, $\rm{\Delta_r}$) parameter space. The measured results are displayed in Figs.~\ref{Fig2}(${\rm b}$)-(${\rm e}$), where the trajectory of the eigenenergy is projected onto the parameter space spanned by the frequency $\omega$ and transmission loss $\xi$ (see the definition in Methods section). The trajectory is traced by varying the argument of the complex parameter $\text{Z}=\rm{\Delta_r}+\mathit{i}\rm{\Omega^2_{p}}$. In this case, we set the system to be in the real part of energy $\omega = \rm{\Delta_r}$ and to measure the observable loss $\xi$ corresponding to the imaginary part of energy. It can be seen that the system has a topological property with a winding number $\mathcal{W}_\lambda=1$, which serves as a topological invariant for the system~\cite{wang2021topological,lee2019topological}, is robust to parameter variations, as evidenced by Figs.~\ref{Fig2}(${\rm c}$)-(${\rm e}$).

At small $\rm{\Omega_p^2}$, dissipation is weak and the system exhibits quasi-Hermitian behavior: the transmission loss remains nearly constant and the spectral path lies close to the real axis, as shown in Fig.~\ref{Fig2}(${\rm b}$). However, as $\rm{\Omega_p^2}$ increases, Rydberg many-body interactions induce significant dissipation, shifting the dynamics into the non-Hermitian regime. Similar to the theory, the spectral trajectory becomes larger as the interaction increases. Thus, the trajectory expands significantly, highlighting the role of interaction-induced dissipation.

The resulting transmission trajectories reveal robust loop structures that are directly tied to the spectral winding number. The absence of such loops in the weak-interaction limit $V\rightarrow0$ underscores the essential role of strong interactions in breaking Hermitian constraints and enabling topology driven by dissipation. These results illustrate a continuous transition mediated by interaction strength, where the system evolves from a Hermitian-like regime with trivial spectral lines into a non-Hermitian phase with non-zero spectral winding.

\begin{figure*}
\centering
\includegraphics[width=1\linewidth]{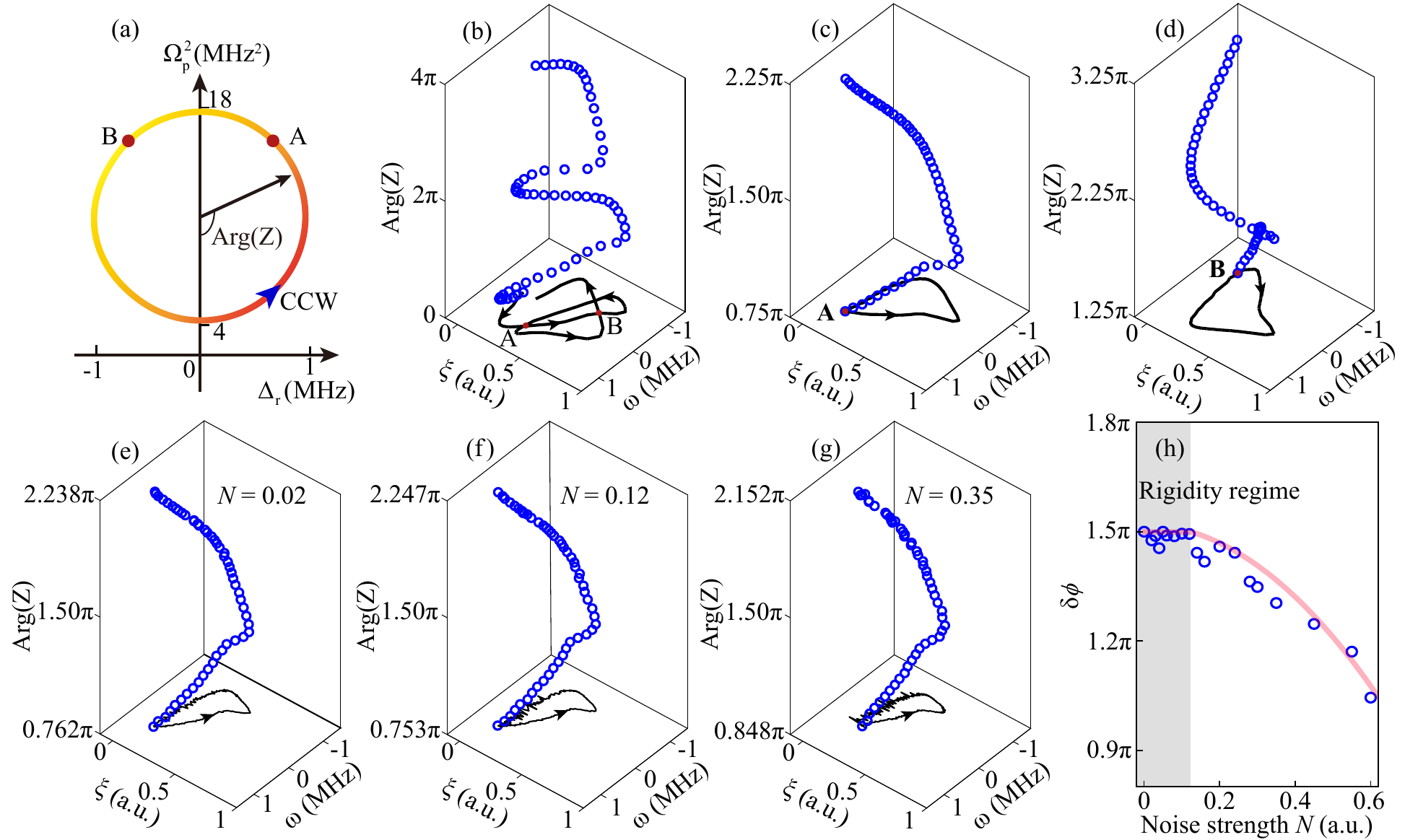}
\caption{\textbf{Fractional winding phase and opposite winding.} (a) The counterclockwise (CCW) scan paths of the parameters $\rm{\Omega_p^2}$ and $\rm{\Delta_r}$ with different starting points A and B. (b) The measured spectral topological trajectory and projection (the black line) under the CCW scan with Arg(Z) ranging from 0 $\sim 4\pi$. (c) and (d) are the measured spectral topological trajectories and projections (the black line) under the CCW scanning path. They correspond to the starting points A and B in panel (a),  with Arg(Z) from $0.75\pi\sim 2.25\pi$ and $1.25\pi/4\sim 3.25\pi$, respectively. The winding number $\mathcal{W}_\lambda$ = 1 for (c) and $\mathcal{W}_\lambda$ = -1 for (d). In panel (c), the phase winding in the complex plane of the parameter space is $1.5\pi$ for a closed loop with $\mathcal{W}_\lambda$ = 1. (e-g) are measured spectral topological trajectories and projections for noise strengths $N$ = 0.02, 0.12, and 0.35, with a fixed $\rm{T_s}=35~\rm{\mu s}$. (h) The accumulated phase $\delta \phi$ for forming a quantized topological structure as a function of $N$. The red line represents the fitted curve using a piecewise function: $y$ = 1.5$\pi$ for $x \leq 0.12$, and $y = c-dx^2$ for $x > 0.12$ (c = 1.52$\pi$ and d = 3.9), and the grey region represents the rigidity regime to $N$.}
\label{Fig5}
\end{figure*}

\subsection*{Twisted topology}
In our system, the direction of parameter scanning governs a fundamental transition between dominant quantum processes: from absorption to interaction-driven dissipation. This is vividly captured in Fig.~\ref{Fig3}(a), where the imaginary part of the coherence, $\rm{Im(\rho_{eg}})$, exhibits a twisted hysteresis between the up-scan and down-scan processes of the probe field $\rm{\Omega_p}$. The crossing of the transmission spectra reveals a dynamical intersection point: at lower $\rm{\Omega_p}$, absorption dominates, as evidenced by the larger $\rm{Im(\rho_{eg}})$ during the increasing scan compared to the decreasing scan. In contrast, at higher $\rm{\Omega_p}$ strong Rydberg interactions induce enhanced dissipation, reversing the hysteresis so that the decreasing scan now yields greater $\rm{Im(\rho_{eg}})$. This crossover reflects the competition between radiative absorption and the non-Hermitian dissipation arising from Rydberg many-body interaction. Thus, the resulting topological winding number in the regimes of absorption-dominated and dissipation-dominated has different signs, as illustrated in Fig.~\ref{Fig3}(b).

The dynamical nature of this competition is further controlled by the scanning time, which directly influences the effective interaction strength between Rydberg atoms. As shown in Figs.~\ref{Fig3}(c)-(h), the projection of the spectral trajectories in the ($\xi$, $\omega$) plane evolves dramatically with the scan period $\rm{T_s}$. When $\rm{T_s}$ is long (e.g., 19 $\mu$s), the system responds adiabatically, and the transmission loop exhibits a well-defined winding structure characteristic of a certain topological phase. As $\rm{T_s}$ shortens, the increasing non-adiabaticity alters the interplay between absorption and dissipation, leading to growing distortion of the loop. Eventually, at sufficiently short periods (e.g., 1 $\mu$s), the loop becomes twisted and even self-intersecting, signaling a topological transition in which the winding number $\mathcal{W}_\lambda$ changes sign. 

This evolution underscores the role of the scanning time as a control parameter for topology: it regulates whether the system can follow the instantaneous eigenstates or is driven into a non-equilibrium regime where dissipation and coherent dynamics compete. The resulting spectral loops are not mere geometric constructs but reflect the accumulation of non-adiabatic phase effects and the non-reciprocal flow of energy and coherence in an open quantum system. Thus, by varying Arg(Z) and $\rm{T_s}$, we directly manipulate the topological structure of the system’s response, demonstrating the rich interplay between non-Hermitian topology, many-body interactions, and quantum dynamics out of equilibrium.

\begin{figure*}
\centering
\includegraphics[width=1\linewidth]{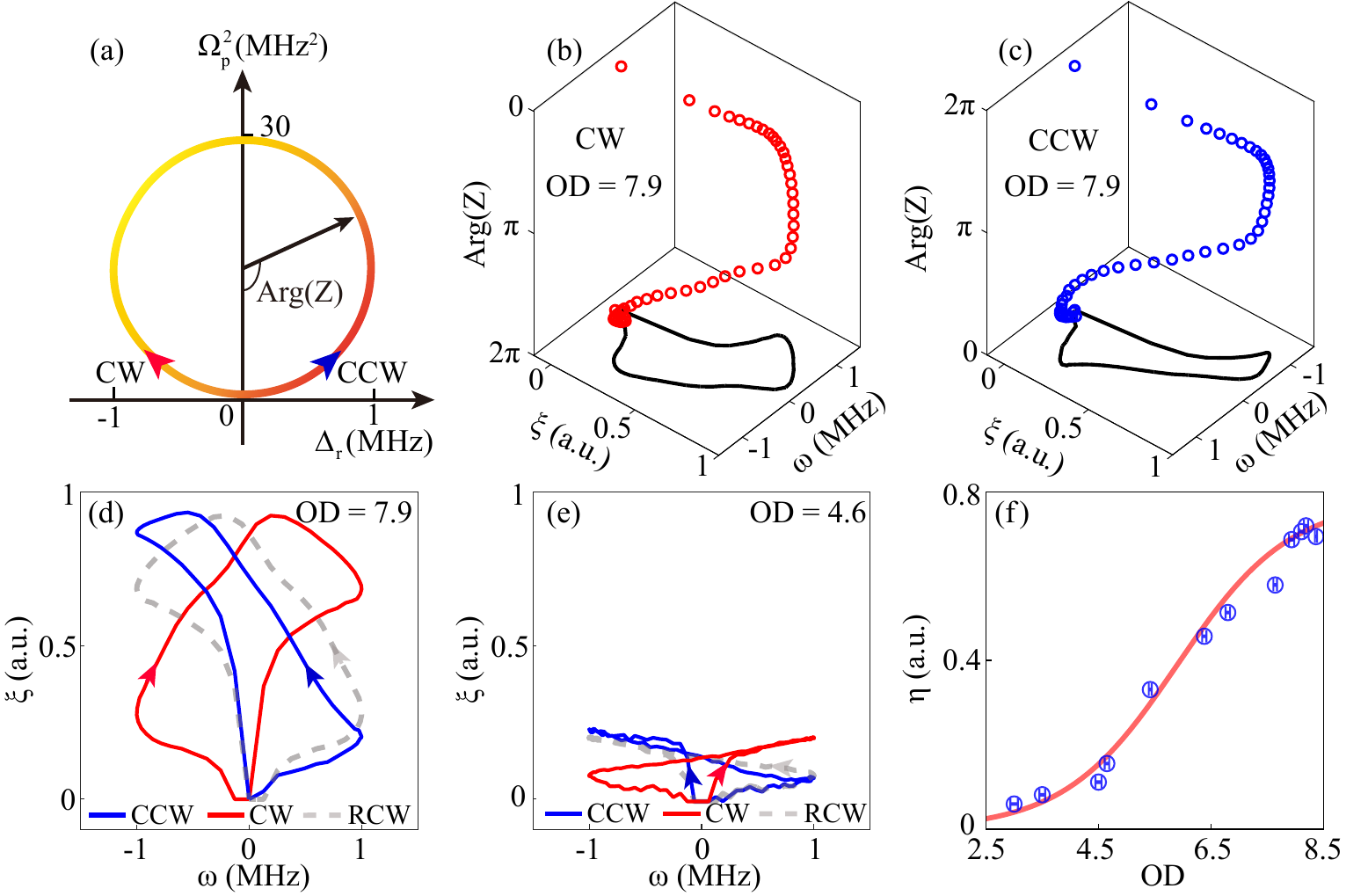}
\caption{\textbf{Chirality symmetry breaking.} (a) The clockwise (CW) and counterclockwise (CCW) scan paths of the parameters $\rm{\Omega_p^2}$ and $\rm{\Delta_r}$, corresponding to the chiral operation. (b) The spectral topological trajectory and projection (the black line) under the CW scanning path. (c) The spectral topological trajectory and projection (the black line) under the CCW scanning path. (d, e) The projections of spectra under two different scanning paths on the parameter plane, at the condition of optical depth OD = 7.9 (d) and OD = 4.6 (e). The grey dash curve denotes the mirror curve of the projected topological trajectory under the CW scanning. (f) The nonreciprocity parameter $\eta$ as a function of OD, and the red curve is a function fit $y = L / [1 + e^{-k(x - x_0)}]$ ($L=0.78$, $k=1.0$, and $x_0=5.9$).}
\label{Fig4}
\end{figure*}

\subsection*{Fractional winding phase}
The controlled cyclic evolution is key to probing the system's initial state-dependent topological properties. Figure~\ref{Fig5}(a) shows two CCW scanning paths of the parameters $\rm{\Omega^2_p}$ and $\rm{\Delta_r}$, originating from distinct starting points A and B. We first change the complex parameter Arg(Z) by tracing the path twice around the origin (from 0 to $4\pi$) to detect the system response data in a larger data range. The measured spectral response, as shown in Fig. \ref{Fig5}(b), reveals a unique topological trajectory in the complex energy plane, in which the trajectory is not enclosed due to the atom's loss throughout the excitation process.

However, when scanning the parameters ($\rm{\Omega^2_p}$, $\rm{\Delta_r}$) starting from either point A or point B in Fig.~\ref{Fig5}(a) or Fig.~\ref{Fig5}(b), a nontrivial spectral topology is clearly observed, as illustrated in Fig.~\ref{Fig5}(c) and Fig.~\ref{Fig5}(d). The winding number $\mathcal{W}_\lambda$ is extracted from such trajectories. For the CCW scan with Arg(Z) evolving from $0.75\pi\sim 2.25\pi$ (a total phase change of $\delta\phi=1.5\pi$, where $\delta\phi$ can be calculated through the conditions $\xi(\phi_1)=\xi(\phi_2)$, and $\omega(\phi_1)=\omega(\phi_2)$), the measured trajectory in Fig.~\ref{Fig5}(c) yields a winding number of $\mathcal{W}_\lambda$ = +1. Conversely, for the path with Arg(Z) from $1.25\pi\sim 3.25\pi$ in Fig.~\ref{Fig5}(d), the trajectory winds in the opposite direction, resulting in $\mathcal{W}_\lambda$ = -1. The sign change of the winding number $\mathcal{W}_\lambda$ signifies a reversal in the direction of phase accumulation since the open quantum many-body system has initial state-dependent non-equilibrium dynamical evolution. 

In Fig.~\ref{Fig5}(c), the observation of a fractional phase winding of $1.5\pi$ within a single closed loop in parameter space exhibits a non-trivial manifestation of the system's underlying non-Hermitian topological structure. In our open quantum many-body system, the interaction-induced dissipation disrupts the adiabatic condition. This prevents the system's state from continuously tracking the instantaneous eigenstates, causing a deviation from the coherent, closed-system behavior. As a result, the geometric phase accumulation is incomplete, resulting in a fractional winding phase instead of a full integer multiple of $2\pi$. This effect highlights how interaction-induced dissipation fundamentally alters the topological response by breaking the adiabatic approximation that underpins quantized phase evolution in Hermitian systems.

Furthermore, we demonstrate the remarkable robustness of the fractional winding phase of $\delta\phi=1.5\pi$ against controlled perturbations, as shown in Figs.~\ref{Fig5}(e)-(g). By introducing white noise of varying strength to the control parameter $\rm{\Omega^2_p}$, we observe that while the topological trajectory in the complex energy plane distorts, the winding number maintains its quantization $\mathcal{W}_\lambda$ = 1 around at $\delta\phi=1.5\pi \pm 0.05\pi$ [see the rigidity regime in Fig.~\ref{Fig5}(h)]. This stability directly manifests the topological protection inherent in the non-Hermitian spectral loop, suggesting that the observed fractional phase is not a fragile artifact but a robust topological invariant resilient to experimental imperfections.

\subsection*{Chirality symmetry breaking}
Figure~\ref{Fig4} demonstrates the emergence of reverse spectral topology controlled by the scanning direction of Arg(Z). As illustrated in Fig.~\ref{Fig4}(a), two distinct scanning paths are implemented in the complex parameter space defined by $\rm{\Omega^2_p}$ and $\rm{\Delta_r}$: a CCW and a CW trajectory. These directions are not equivalent due to the dissipative, non-conservative nature of the system, which breaks time-reversal symmetry at the level of the effective non-Hermitian Hamiltonian.

Figure~\ref{Fig4} (b) and (c) show the experimentally measured spectral responses under these two scanning directions, in which the topological trajectory in the complex energy plane and its projection (black curve) are mapped. The trajectory in Fig.~\ref{Fig4}(b), corresponding to CW scanning, exhibits a loop structure with characteristic winding, while the reversed scan in Fig.~\ref{Fig4}(c) results in a distinct, mirror-like path, highlighting the directional sensitivity of the spectral topology. This demonstrates that the spectral topology is not a static property but is intrinsically tied to the direction of parameter variation, revealing the dynamical and geometry-dependent nature of non-Hermitian energy spectra.

Figure~\ref{Fig4}(d) directly compares the projections of these spectral trajectories. The clear separation between the two curves underscores the non-reciprocity and path-dependence of the system's response, revealing a signature of broken chirality symmetry. This behavior arises from the combination of strong Rydberg-mediated dissipation and the adiabatic (or non-adiabatic) driving of system parameters, which together break chiral symmetry in the measurement trajectory. These results provide a direct visualisation of how dissipation and interactions collaborate to produce directional-sensitive topological phases in open quantum systems.

To quantify the property of the chirality symmetry breaking of the system~\cite{miao2024spontaneous}, we define the nonreciprocity parameter 
\begin{equation}
\eta=\oint_{C_\mathrm{Z}} \mathrm{d} \vec{\mathrm{Z}}~{{\left| \mathit{O}_{\mathrm{RCW}}(\rm{Z}) - \mathit{O}_{\mathrm{CCW}}(\rm{Z})\right|}}.
\end{equation}
Here, $O_{\mathrm{RCW}}(\rm{Z})$ and $O_{\mathrm{CCW}}(\rm{Z})$ denote the measured observables along the mirror-symmetric clockwise (RCW) and CCW scanning paths at the complex parameter Z. The parameter $\eta$ quantifies the relative asymmetry between the system’s responses under these two scanning directions. The non-zero value indicates a signature of chirality symmetry breaking as it measures the system's asymmetry under the chiral operation of reversing the parameter scan direction. In the experiment, we evaluate an effective nonreciprocity parameter by taking the observable $O(\rm{Z})$ as the probe transmission loss $\xi$. 

To quantitatively investigate the chirality symmetry breaking, we systematically vary the optical density (OD) from 3.0 to 8.3. As the OD increases, the separation between the RCW and CCW projected trajectories becomes more distinct. This trend arises because a higher OD involves more atoms in the interaction volume, which enhances the effective Rydberg-mediated interaction-induced dissipation $\gamma_\mathrm{eff}$ in the system. The increased dissipation further reinforces the non-Hermitian topological character, resulting in a more clearly defined spectral winding loop, as clearly seen at the two cases of OD = 7.9 in Fig.~\ref{Fig4}(d) and OD = 4.6 in Fig.~\ref{Fig4}(e). This direct relationship is conclusively demonstrated in Fig.~\ref{Fig4}(f), where the nonreciprocity parameter $\eta$ increases against with OD. 

\subsection*{Conclusion}
In this work, we have demonstrated an experiment to study non-Hermitian topological physics in the presence of strong interactions and dissipation in cold Rydberg atomic gases. Through precise manipulation of system parameters, such as probe intensity, two-photon detuning, and scanning time, we have shown the emergence of spectral loops in the complex energy plane, indicating a transition from Hermitian to non-Hermitian regimes. The observed twisting and self-intersection of these loops under varying scanning times reveal a dynamical topological phase transition, accompanied by a sign change in the winding number. By preparing the system in different initial states, we also accessed a fractional phase within a closed winding loop, highlighting the role of the initial state in accessing exotic non-Hermitian topological features. Furthermore, the nonreciprocal spectral responses under CW and CCW parameter scans provide direct evidence of broken chirality symmetry, an intrinsic characteristic of non-Hermitian dynamical topology. These findings underscore the rich interplay between dissipation, interactions, and topology in open quantum systems, and establish cold Rydberg gases as a platform for studying dynamical topological phenomena beyond the Hermitian framework. 

The long-range interactions between Rydberg atoms play a vital role, as they not only enhance many-body correlations but also mediate the dissipative processes that are essential for realizing non-Hermitian topological states. This results in non-static topological properties that evolve dynamically with the scanning process. The interplay between long-range interactions and dissipation enriches the topological landscape, enabling the emergence of phases that are otherwise inaccessible in short-range interacting systems. The observed topology is not merely a static property of the Hamiltonian but emerges from the interplay between coherent dynamics, dissipation, and the measurement process itself, revealing the non-equilibrium nature of topology in open quantum systems. This work opens new avenues for exploring non-equilibrium topological phases and their potential applications in quantum simulation and information processing.

\section*{Methods}
\subsection*{Details of the experimental setup}
Our investigation into non-Hermitian spectral topology utilizes a cold ensemble of $^{85}$Rb atoms confined in a three-dimensional magneto-optic trap. Through optical pumping, the atoms are prepared in the hyperfine ground state $\ket{g} = \ket{5S_{1/2}, F = 3}$. To ensure a stable magnetic environment and minimize dephasing, the MOT is enclosed within a double-layer magnetic shield, which suppresses the residual magnetic field to below 10 mGauss. This configuration can avoid the dephasing from the earth's magnetic field. The atoms are excited to the Rydberg state $\ket{r} = \ket{40D_{5/2}}$ via a two-photon transition. A probe beam ($\omega_{p} \approx$ 20 $\mu$m) drives the $\ket{g} \rightarrow \ket{e} = \ket{5P_{3/2}, F = 4}$ transition, while a counter-propagating coupling beam ($\omega_{c} \approx$ 30 $\mu$m) drives the $\ket{e} \rightarrow \ket{r}$ transition. To ensure the high coherence and frequency stability required for EIT, probe and coupling lasers are actively locked using the Pound-Drever-Hall method. The system's response is characterized by measuring the probe beam's transmission through the atomic ensemble using a photo-multiplier tube. 

In experiment, a sine wave and a cosine wave generated by the signal generator (RIGOL DG4062) were applied to the acousto-optic modulator (AOM). Specifically, a cosine wave in the form of $\text{cos}(2\pi t/{\rm{T_s}}+\theta_1)+C$ was applied for amplitude modulation (AM) on the AOM to control $\rm\Omega_p^2$, while a cosine wave in the form of $ \text{sin}(2\pi t/{\rm{T_s}} +\theta_2)+D$ was utilized for frequency modulation (FM) of the AOM to control $\rm\Delta_p$. This configuration generates the variable $\rm{Arg(Z)}$ in the parameter space, where $\text{Z}=\rm{\Delta_r}+i\rm{\Omega_{p}^2}$, as illustrated in Fig.~\ref{Fig1}(c). Here, $\rm{T_s}$ denotes the scanning period, $\theta_1$ and $\theta_2$ is the initial phase, C and D are constant offsets. The initial phases $\theta_1$ and $\theta_2$, determine the rotation direction of the parameter $\rm{Arg(Z)}$: a CW rotation corresponds to $\theta_1=\theta_2$, while a CCW rotation is achieved when $|\theta_2-\theta_1|=\pi$. 

\subsection*{Lindblad master equation}
We simulate transmission and absorption in the Rydberg atomic dissipation system by solving the Lindblad master equation, which is written as
\begin{align}
{\dot \rho _{gg}}&=i \frac{\rm{\Omega_p}}{2} (\rho_{eg} - \rho_{ge}) + \gamma_1 \rho_{ee},\\
{\dot \rho _{ee}}&=i \frac{\rm{\Omega_p}}{2} (\rho_{ge} - \rho_{eg}) + i \frac{\rm{\Omega_c}}{2} (\rho_{re} - \rho_{er})- \gamma_1 \rho_{ee} + \gamma_2 \rho_{rr}, \\
{\dot \rho _{rr}}&=i \frac{\rm{\Omega_c}}{2} (\rho_{er} - \rho_{re})- \gamma_2 \rho_{rr},\\
{\dot \rho _{er}}&=i \frac{\rm{\Omega_c}}{2} (\rho_{rr} - \rho_{ee}) + i \frac{\rm{\Omega_p}}{2} \rho_{gr} + \left(i \rm{\Delta _c} - \frac{\gamma_2}{2} - \frac{\gamma_1}{2}\right) \rho_{er},\\ 
{\dot \rho _{gr}}&= i \frac{\rm{\Omega_p}}{2} \rho_{er}-i \frac{\rm{\Omega_c}}{2} \rho_{ge} +\left(i (\rm{\Delta _p} -\rm{\Delta _c})-\frac{\gamma_2}{2}\right) \rho_{gr},\\
{\dot \rho _{ge}}&=i \frac{\rm{\Omega_p}}{2} (\rho_{ee} - \rho_{gg}) - i \frac{\rm{\Omega_c}}{2} \rho_{gr} + \left(i \rm{\Delta _p} - \frac{\gamma_1}{2}\right) \rho_{ge}, 
\end{align}
The remaining equations are given by ${\rho _{eg}} = \rho _{ge}^*$, ${\rho _{rg}} = \rho _{gr}^*$, and ${\rho _{re}} = \rho _{er}^*$. In the mean-field approximation ($\gamma_2\rightarrow\gamma_\mathrm{eff}=\gamma+V\rho_{rr}$) \cite{zhang2025exceptional}, we replace the time-dependent population $\rho_{rr}(t)$ by the term $\dot \rho _{rr}(t)$ according to Eq. (5): $ \rho_{rr}(t)=i {\rm{\Omega_c}}/{2\Gamma_2} (\rho_{er}(t) - \rho_{re}(t))-{\dot \rho _{rr}(t)}/{\Gamma_2}$. Thus, the slope of Rydberg population $\dot \rho _{rr}(t)$ plays a crucial role in influencing the transient behavior of $\rho_{rr}(t)$. By solving the time-dependent Lindblad master equation outlined above, we obtained the matrix element $\rho_{eg}$. The theoretical results are presented in Fig.~\ref{Fig3}(a), where we present the dynamic evolution curve of the system as the Rabi frequency $\rm{\Omega_p}$ varies. 

In the experiment, we obtain Loss of the probe field $\xi= 1-I/I_0= 1- e^{-k \operatorname{Im}[\chi] l}$, where $\chi=2m|\mu_{eg}|^2 \rho_{eg}/\left[\varepsilon_0 \rm{\Omega_p}\right]$ is polarizability, $l$ is the distance the probe laser beam travels through the cold atomic gases, $k$ is the wave number of the probe light in vacuum, $m$ is the atomic number density, $\mu_{eg}$ is transition dipole moment, and $I$ and $I_{0}$ represent the intensity of the input and output of the probe light respectively. Under the weak probe approximation, $\xi \propto \operatorname{Im}\left[\rho_{e g}\right] $~\cite{fleischhauer2005electromagnetically,ding2023ergodicity}. Under the condition of weak excitation approximation ($\rm{\Omega_p} \ll \rm{\Omega_c}$ and $\rho_{g g} \approx 1$ ), we can obtain the steady-state solution of the $\rho_{g e}$,
\begin{equation}
\begin{aligned}
\rho_{eg} \simeq & \frac{1}{{\rm{\Delta}_p}^2+\gamma_{1}^2 / 4}\left(\frac{{\rm{\Delta}_p} \rm{\Omega_p}}{2}-\frac{\left|\rm{\Omega_c}\right|^2 \gamma_{1}}{4 \rm{\Omega_p}} \rho_{r r}\right) \\
                  & - \frac{i}{{\rm{\Delta}_p}^2+\gamma_{1}^2 / 4}\left(\frac{\gamma_{1} \rm{\Omega_p}}{4}+\frac{\left|\rm{\Omega_c}\right|^2 {\rm{\Delta}_p}}{2 \rm{\Omega_p}} \rho_{r r}\right).
\end{aligned}
\end{equation}
It is evident that $\mathrm{Im}[\rho_{eg}]$ is proportional to $\rho_{rr}$. Since the interaction-induced dissipation $\gamma_\mathrm{eff}$ opens an additional dissipative channel that depletes the population in $\rho_{rr}$, thereby resulting in the observed non-linear loss of probe transmission.

\section*{Acknowledgements}
We thank Prof. Jin-Shi Xu, Prof. Xiao-Ye Xu and Prof. Xi-Wang Luo for helpful discussions. We acknowledge funding from the National Key R and D Program of China (Grant No. 2022YFA1404002), the National Natural Science Foundation of China (Grant Nos. T2495253, 62435018).

\hspace*{\fill}

\section*{Data Availability}
All experimental data used in this study are available from the corresponding author upon request.

\section*{Author contributions statement}
D.-S.D. conceived the idea and supported this research. J.Z, Y.J.W and S.-Y. S conducted the physical experiments. Y.J.W developed the theoretical model. The manuscript was written by D.-S.D, J.Z., and Y.J.W.  All authors contributed to discussions regarding the results and the analysis contained in the manuscript.

\section*{Competing interests}
The authors declare no competing interests.

\end{document}